\documentclass[journal]{IEEEtran}
\usepackage{color}
\usepackage{graphicx}
\usepackage{epstopdf}
\usepackage{amsmath}
\usepackage{array}

\ifCLASSOPTIONcompsoc
  \usepackage[caption=false,font=normalsize,labelfont=sf,textfont=sf]{subfig}
\else
  \usepackage[caption=false,font=footnotesize]{subfig}
\fi
\begin{document}
\title{\LARGE{An Efficient Deep Learning Model for Automatic Modulation Recognition Based on Parameter Estimation and Transformation}}

\author{Fuxin~Zhang, 
        Chunbo~Luo,~\IEEEmembership{Member,~IEEE,}
        Jialang~Xu,
        and~Yang~Luo 
\thanks{Manuscript received xxx; revised xxx.\textit{(Corresponding authors: Chunbo Luo; Yang Luo.)}}
\thanks{Fuxin~Zhang and Jialang~Xu are with the School of Information and Communication Engineering, University of Electronic Science and Technology of China, Chengdu 611731, China. (e-mail: fxzhang@std.uestc.edu.cn, xujialang@std.uestc.edu.cn)}
\thanks{Chunbo~Luo and Yang~Luo are with the School of Information and Communication Engineering, University of Electronic Science and Technology of China, Chengdu 611731, China, and also with the Yangtze Delta Region Institute (Huzhou), University of Electronic Science and Technology of China, Huzhou 313001, China.  (e-mail: c.luo@uestc.edu.cn, luoyang@uestc.edu.cn)}
}
\maketitle

\begin{abstract}
Automatic modulation recognition (AMR) is a promising technology for intelligent communication receivers to detect signal modulation schemes. Recently, the emerging deep learning (DL) research has facilitated high-performance DL-AMR approaches. However, most DL-AMR models only focus on recognition accuracy, leading to huge model sizes and high computational complexity, while some lightweight and low-complexity models struggle to meet the accuracy requirements. This letter proposes an efficient DL-AMR model based on phase parameter estimation and transformation, with convolutional neural network (CNN) and gated recurrent unit (GRU) as the feature extraction layers, which can achieve high recognition accuracy equivalent to the existing state-of-the-art models but reduces more than a third of the volume of their parameters. Meanwhile, our model is more competitive in training time and test time than the benchmark models with similar recognition accuracy. Moreover, we further propose to compress our model by pruning, which maintains the recognition accuracy higher than 90\% while has less than 1/8 of the number of parameters comparing with state-of-the-art models. 

\end{abstract}


%
\IEEEpeerreviewmaketitle

\section{Introduction}
\IEEEPARstart{A}{utomatic} modulation recognition (AMR) has made it possible to identify signal modulation schemes automatically by the receiver in the non-cooperative communications scenarios, which has various civilian and military applications, such as spectral interference detection, spectrum sensing, cognitive radio, etc \cite{dobre2007survey}. Deep learning (DL) based AMR methods are showing improved performance in terms of recognition accuracy and complexity compared with traditional likelihood-based methods and feature-based methods \cite{o2016convolutional}. 

Prior DL-AMR models have achieved benchmark performance for AMR by using various neural network layers, e.g. convolutional neural network (CNN) \cite{o2016convolutional,hermawan2020cnn,xu2020identification}, recurrent neural network (RNN) \cite{rajendran2018deep,hong2017automatic}, and hybrid network \cite{xu2020spatiotemporal}, etc. Considering simple neural networks lacking the ability to eliminate signal distortion caused by the wireless channel, \cite{yashashwi2018learnable} designed a signal distortion correction module to equalize the carrier frequency and phase offset, which demonstrated the potential to improve DL-AMR models by incorporating expert domain knowledge. With the rapid development of 5G/B5G in recent years, the growth of massive Internet-of-things (IoT) devices demand improved communications performance with limited available resources \cite{shafi20175g}, and thus efficient AMR models are crucially important for the future IoT devices with limited computing and energy resources. Consequently, researchers begin to research the lightweight and low-complexity DL-AMR models by reducing the model size or accelerating the computation time while ensuring the recognition accuracy  \cite{wang2020lightamc,huynh2020mcnet,tunze2020sparsely}, making them increasingly possible to deploy in resource-limited devices. However, the existing models with high recognition accuracy rarely considered the model size and complexity in the design process, while lightweight and low-complexity models struggle to achieve high accuracy. 

In this letter, we propose an efficient DL-AMR model inspired by radio transformer networks (RTN) \cite{o2017introduction}, CNN, and GRU. The original data is processed by a parameter estimation network and a transformation module, and then the spatial and temporal features of the signals are extracted by CNN and gated recurrent unit (GRU) for classification. The model can achieve high recognition accuracy equivalent to state-of-the-art models but with much fewer parameters, while the training time and test time of our model outperform the benchmark models without suffering recognition accuracy. A network pruning method \cite{zhu2017prune} is further applied to compress the model size while maintaining high recognition accuracy, making it a promising candidate for resource-limited systems. 

The contributions of this letter are summarized as follows:
\begin{itemize}
\item An efficient model that can achieve state-of-the-art recognition accuracy over three benchmark datasets but with least parameters is proposed based on parameter estimator, parameter transformer, CNN, and GRU.
\item To efficiently utilize the spatial-temporal features of AMR signals, we propose to decrease the kernel size and feature maps in the CNN layers, and introduce parameter estimator and transformer to reduce the adverse effects on phase, leading to improved recognition accuracy. 
\item We demonstrate that the proposed lightweight model can be further compressed by five times using pruning method to fit the scenarios with extremely limited resources. 
\end{itemize}
\section{Signal Model and Proposed System Model}
\label{section1}
\subsection{Signal Model}
After the signal passes through the channel and is sampled, the equivalent baseband signal can be expressed by:
\begin{equation}
y[l]=A[l]e^{j(\omega l+\varphi)} x[l]+n[l], l=1, \ldots, L,
\end{equation}
where $x[l]$ is the signal modulated by the transmitter in a certain modulation scheme, $n[l]$ denotes the complex Additive Gaussian Noise (AWGN), $A[l]$ represents the channel gain, $\omega$ is the frequency offset, $\varphi$ is the phase offset, $y[l]$ denotes the $l$-th value observed by the receiver and $L$ is the number of symbols in a signal sample. To facilitate data processing and modulation recognition, the received signals can be stored in in-phase/quadrature (I/Q) form, denoted as $\textbf{y} = [\Re\{y[1]\}, ..., \Re\{y[L]\}; \Im\{y[1]\}, ..., \Im\{y[L]\}]$.
\subsection{The Proposed DL Model}
\begin{figure}[htbp]
\centering
\includegraphics[scale=.3]{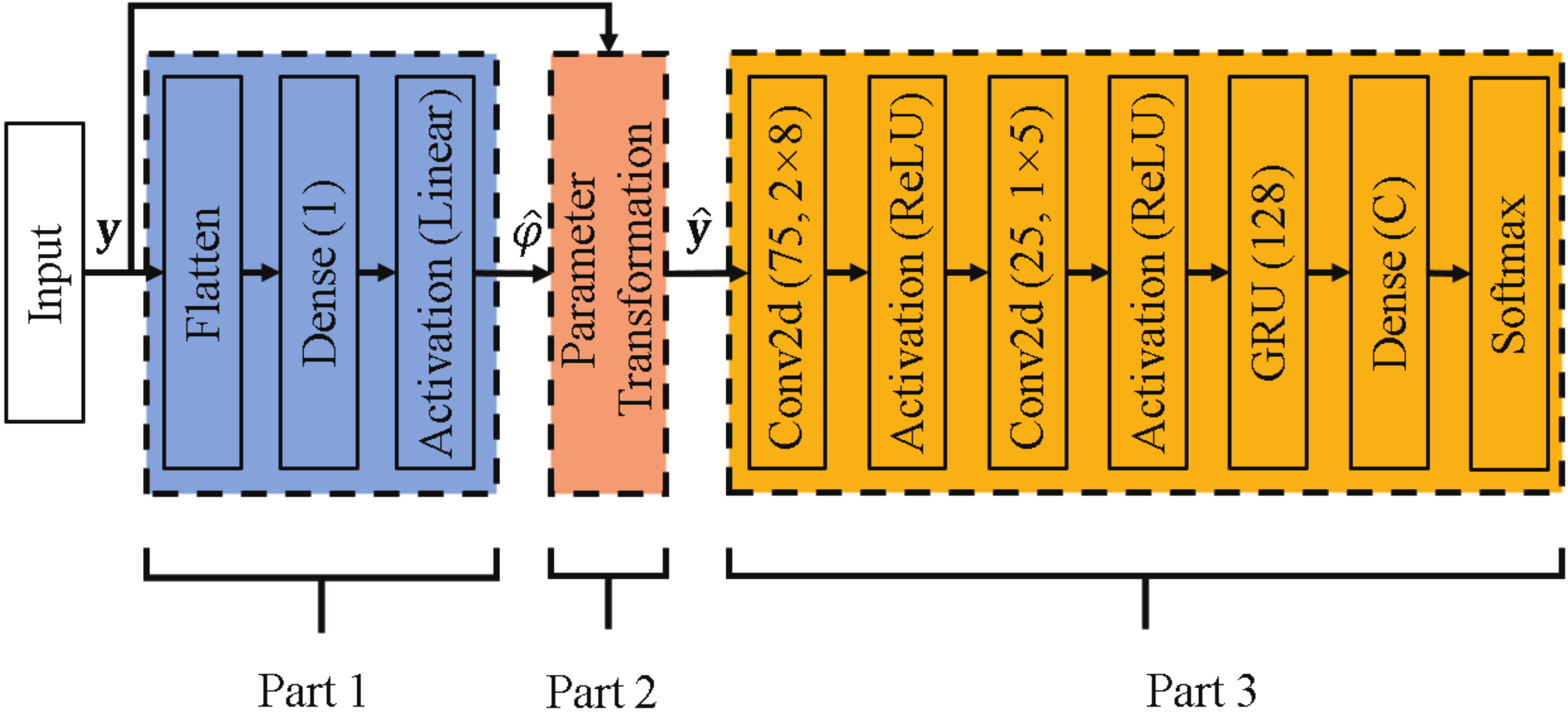}
\caption{The structure of the proposed PET-CGDNN.}
\label{fig1}
\end{figure}

The proposed parameter estimation and transformation based CNN-GRU deep neural network (PET-CGDNN) model comprises a parameter estimator, a parameter transformer, and a hybrid neural network, which is shown in Fig. \ref{fig1}. 

As described in the signal model, the sampled I/Q data is affected by noise and interference from the channel and imperfect hardware design, which may result in adverse effects such as temporal shifting, linear mixing/rotating, and spinning of the received signal. Many of such effects can be inverted using parametric transformations according to classic signal processing theories. Hence, the parameter estimator (Part 1) and the parameter transformer (Part 2) have been introduced in our model to extract phase offset related information and perform phase parameter transformation, which are the key to enhancing the recognition accuracy of our model. 

The parameter estimator in Part 1 can estimate the phase parameter by co-training with the subsequent model. For each I/Q sample with the dimension of 2 × $L$ ($L$ = 128 or 1024), one phase parameter, which carries the phase offset related information, is estimated by a neural network composed of a Flatten layer and a Dense layer (fully connected layer). The input signal $\textbf{y}$ is flattened into a vector by the Flatten layer to satisfy the input dimension of the Dense layer, and then the data in this vector are correlated through the Dense layer to obtain the phase parameter $\hat{\varphi}$. The activation function of the Dense layer is Linear, which yields an estimated phase parameter from a continuous unbounded range.

The parameter transformer in Part 2 is a customized layer, which performs parametric inverse transformation by taking the input $\textbf{y}$ and $\hat{\varphi}$, given by:
\begin{equation}\hat{y}[l] = y[l] e^{-j \hat{\varphi}} = \left[\begin{array}{l}
\Re\{y[l]\} \cos \hat{\varphi}+\Im\{y[l]\} \sin \hat{\varphi} \\
\Im\{y[l]\} \cos \hat{\varphi}-\Re\{y[l]\}\sin \hat{\varphi}
\end{array}\right],
\end{equation}
where $\hat{\varphi}$ is the estimated phase parameter and ${\hat{\textbf{y}}} = [\hat{y}[1], ..., \hat{y}[L]]$ is the output of Part 2. 

Part 3 consists of CNN, GRU, and Dense layer which can realize feature extraction and classification. The first convolutional layer has 75 filters and 2 × 8 kernel size, which extracts the spatial features of the signal, while the second convolutional layer has 25 filters and 1 × 5 kernel size, which further compacts the extracted features. The subsequent GRU layer extracts the temporal features of the signal with 128 units. Finally, the classification task is completed through the Dense layer, and the number of hidden units is C, which is equal to the number of modulation classes. The first two convolution layers use rectified linear unit (ReLU) activation functions, and the activation function of the last Dense layer is Softmax. Aiming at building a model with a smaller size and low computational cost, the corresponding network in the third part is designed to control the model size with a small number of parameters in the CNN layer.


\subsection{Model Pruning Method}
\label{sub pruning}
Although our model already has a small number of parameters, CNN based structures usually have intrinsic redundancy, which could cause a large model size and unnecessary computational cost. To further compress the model, a network pruning method is adopted to reduce the redundancy. We apply the following approach for the proposed DL-AMR model, aiming to maintain a high recognition accuracy while reducing the model size, modeled as follows,
\begin{equation} \label{eqn2}
  \begin{split}
  s_{t} = s_{f} + \left(s_{i} - s_{f}\right)\left(1 - (t - t_{0})/{n \Delta t}\right)^{3},
  \end{split}
\end{equation}
where $s_{t}$ is initial sparsity value, $s_{f}$ denotes final sparsity value, $t_{0}$ refers to the step to start training with pruning frequency $\Delta t$, the fine-tuning process is divided into $n$ steps to gradually increase the sparsity, and $t\in\{t_{0}, t_{0}+\Delta t, \ldots, t_{0}+n \Delta t\}$. Binary mask variables of the same size and shape as the weights tensor are added to Dense, CNN, and GRU layers. The weights masks of each layer are sorted in the fine-tuning process, and the corresponding weights in the smallest portion are set to zero until reaching the target sparsity goal.

\section{Datasets and Implementation Details}
\label{section2}
The experiments are conducted on RML2016.10a, RML2016.10b \cite{o2016radio} and RML2018.01a \cite{o2018over} datasets with the input data dimension of 2 × 128, 2 × 128 and 2 × 1024, respectively. RML2016.10a dataset includes 220,000 modulated signals with 11 commonly used modulation schemes, RML2016.10b dataset contains 1,200,000 signals with 10 schemes, and RML2018.01a dataset has over 2.5 million signals with 24 modulation schemes. Owing to the hardware limitations, only half of RML2018.01a dataset is randomly selected in our experiment. RML2016.10a and RML2016.10b are generated by simulation using GNU radio while RML2018.01a is produced in a laboratory environment. 

We divide the datasets into training, validation and test at the ratio of 6:2:2 per class with random selection. The loss function is categorical cross-entropy, and the optimizer is Adam. When the validation loss does not decrease in 5 epochs, it is multiplied by a coefficient of 0.5. When the validation loss does not decrease in 50 epochs, the training process is stopped and the trained model is saved with the minimum validation loss. The experiments are implemented using GeForce GTX 1080Ti GPU and Keras with Tensorflow as the backend.

Key AMR models are implemented to provide benchmark comparison which includes IC-AMCNET \cite{hermawan2020cnn}, MCNET \cite{huynh2020mcnet}, LSTM2 \cite{rajendran2018deep}, GRU2 \cite{hong2017automatic}, MCLDNN \cite{xu2020spatiotemporal}.

\section{Experimental Results and Discussion}
\label{section3}
\subsection{Model Performance Measurement}

Several indicators are selected in Table \ref{table1} for performance comparison and complexity analysis, including the number of parameters, training time, test time, highest accuracy of all SNRs (signal-to-noise ratio), and average accuracy of all SNRs. All models are assessed on the three datasets, with input and output layers adjusted to fit the data dimensions, so the number of parameters varies, as shown in the table. Our models in Table \ref{table1} are the vanilla version without pruning.
\begin{table}[htbp]
\centering
\caption{Model comparison on three datasets (A: RML2016.10a, B: RML2016.10b, C: RML2018.01a).}
\resizebox{0.48\textwidth}{!}{
\begin{tabular}{|c|c|c|c|c|c|c|}
\hline
Model& Datasets& Parameters& 
\begin{tabular}[c]{@{}c@{}}Training time\\ (second/epoch)\end{tabular} & \begin{tabular}[c]{@{}c@{}}Test time\\ (ms/sample)\end{tabular} & \begin{tabular}[c]{@{}c@{}}Highest\\ accuracy\end{tabular} &\begin{tabular}[c]{@{}c@{}}Average\\ accuracy\end{tabular}  \\ \hline
IC-AMCNET & \begin{tabular}[c]{@{}c@{}}A\\B\\C\end{tabular}& \begin{tabular}[c]{@{}c@{}}1,264,011\\ 1,263,882\\ 8,605,720\end{tabular} & \begin{tabular}[c]{@{}c@{}}6\\ 34\\ 172\end{tabular}         & \begin{tabular}[c]{@{}c@{}}\textbf{0.036}\\\textbf{0.015}\\0.063\end{tabular} &\begin{tabular}[c]{@{}c@{}}85.59\%\\ 92.82\%\\ 94.42\%\end{tabular} &\begin{tabular}[c]{@{}c@{}}56.83\%\\ 62.15\%\\ 58.4\%\end{tabular}\\ \hline
MCNET& \begin{tabular}[c]{@{}c@{}}A\\ B\\ C\end{tabular} &  \begin{tabular}[c]{@{}c@{}}121,511\\ 121,226\\ 126,616\end{tabular}    & \begin{tabular}[c]{@{}c@{}}8\\ 46\\ \textbf{100}\end{tabular}                  & \begin{tabular}[c]{@{}c@{}}0.041\\ 0.017\\\textbf{0.053}\end{tabular}   & \begin{tabular}[c]{@{}c@{}}83.91\%\\ 89.08\%\\ 91.21\%\end{tabular} &\begin{tabular}[c]{@{}c@{}}56.63\%\\ 60.95\%\\ 57.04\%\end{tabular}\\ \hline
LSTM2 & \begin{tabular}[c]{@{}c@{}}A\\ B\\ C\end{tabular} &  \begin{tabular}[c]{@{}c@{}}201,099\\ 200,970\\ 202,776\end{tabular}    & \begin{tabular}[c]{@{}c@{}}11\\ 56\\ 497\end{tabular}                  & \begin{tabular}[c]{@{}c@{}}0.047\\ 0.032\\ 0.239\end{tabular}   & \begin{tabular}[c]{@{}c@{}}91.41\%\\ \textbf{94.01\%}\\ \textbf{98.39\%}\end{tabular} &\begin{tabular}[c]{@{}c@{}}60.56\%\\ 63.28\%\\ \textbf{65.49}\%\end{tabular}\\ \hline
GRU2& \begin{tabular}[c]{@{}c@{}}A\\ B\\ C\end{tabular} &  \begin{tabular}[c]{@{}c@{}}151,179\\ 151,050\\ 152,856\end{tabular}   & \begin{tabular}[c]{@{}c@{}}9\\ 48\\ 313\end{tabular}                   & \begin{tabular}[c]{@{}c@{}}0.043\\ 0.026\\ 0.151\end{tabular}   & \begin{tabular}[c]{@{}c@{}}87.86\%\\ 93.63\%\\ 98.24\%\end{tabular} &\begin{tabular}[c]{@{}c@{}}58.9\%\\ 63.7\%\\ 63.17\%\end{tabular}\\ \hline
MCLDNN& \begin{tabular}[c]{@{}c@{}}A\\ B\\ C\end{tabular} &  \begin{tabular}[c]{@{}c@{}}406,199\\ 406,070\\ 407,876\end{tabular}    & \begin{tabular}[c]{@{}c@{}}17\\ 90\\ 662\end{tabular}                  & \begin{tabular}[c]{@{}c@{}}0.061\\ 0.045\\ 0.296\end{tabular}   & \begin{tabular}[c]{@{}c@{}}\textbf{92.95\%}\\ 93.86\%\\ 98.11\%\end{tabular} &\begin{tabular}[c]{@{}c@{}}\textbf{62.08}\%\\ 63.78\%\\ 62.59\%\end{tabular}\\ \hline
\begin{tabular}[c]{@{}c@{}}PET-CGDNN\\ (Ours)\end{tabular} &\begin{tabular}[c]{@{}c@{}}A\\ B\\ C\end{tabular} & \begin{tabular}[c]{@{}c@{}}\textbf{71,871}\\ \textbf{71,742}\\ \textbf{75,340}\end{tabular}                              & \begin{tabular}[c]{@{}c@{}}\textbf{6}\\ \textbf{33}\\ 208\end{tabular}                   & \begin{tabular}[c]{@{}c@{}}0.039\\ 0.02\\ 0.099\end{tabular}    & \begin{tabular}[c]{@{}c@{}}91.36\%\\ 93.41\%\\ 98.16\%\end{tabular} &\begin{tabular}[c]{@{}c@{}}60.44\%\\ \textbf{63.82}\%\\ 63.01\%\end{tabular}\\ \hline
\end{tabular}}
\label{table1}
\end{table}

It is clear that the proposed PET-CGDNN model without pruning is already with the least parameters compared with benchmark models in Table \ref{table1}. The time cost of PET-CGDNN has obvious advantages compared with GRU2, LSTM2, and MCLDNN, and is comparable with IC-ACMNET and MCNET but having much higher accuracy. 

Note that the parameters of PET-CGDNN are relatively stable when tested on datasets with a larger input dimension. The IC-AMCNET, originally designed for the dataset with an input data length of 128, would see 6.8 times more parameters for higher-dimensional input data (RML2018.01a). In addition, IC-AMCNET and MCNET are composed of CNNs, which have lower computational complexity than RNN, leading to a low level of training and test time. PET-CGDNN has some disadvantages in time cost compared with IC-AMCNET and MCNET, but the relatively higher cost is offset by the higher recognition accuracy in all test cases. Specifically, PET-CGDNN has the shortest training time on RML2016.10b. Comparing with benchmark high accuracy models such as MCLDNN and LSTM, our model sacrifices little recognition accuracy but greatly reduces the complexity and model size. 
\begin{figure}[htbp]
\centering
\subfloat[]{\includegraphics[width=0.5\columnwidth]{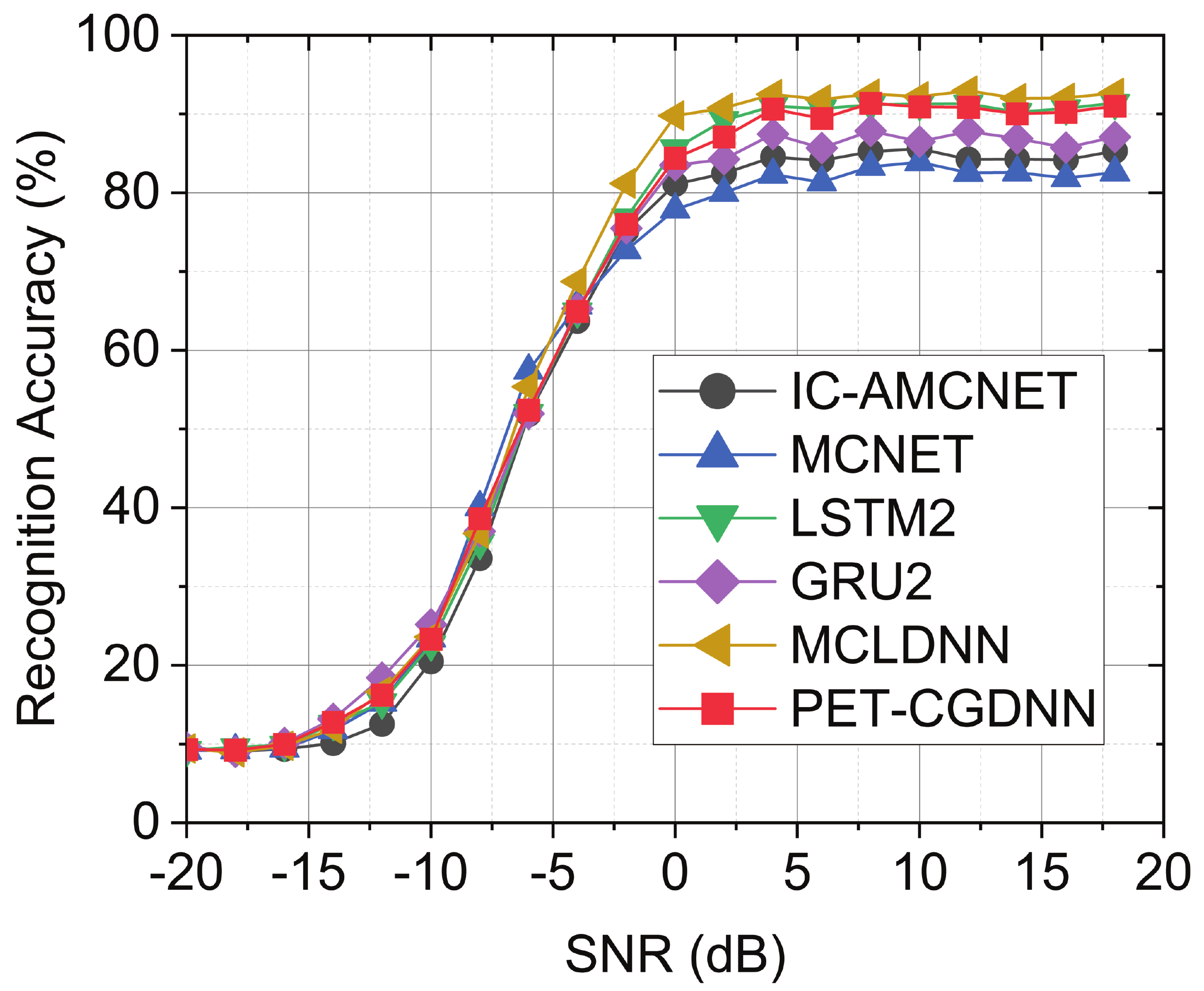}
\label{fig16a}}
\subfloat[]{\includegraphics[width=0.5\columnwidth]{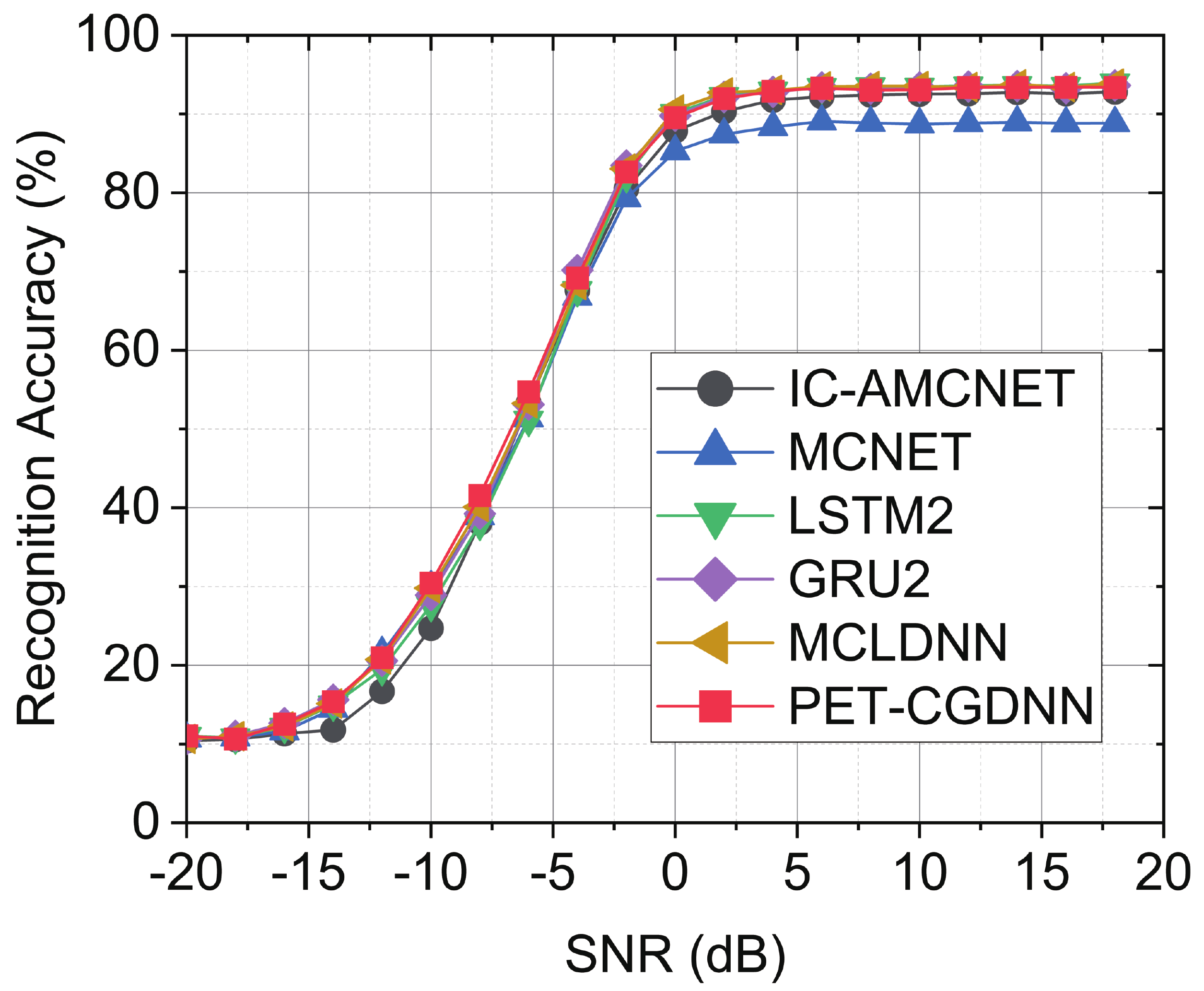}
\label{fig16b}}
\\
\subfloat[]{\includegraphics[width=0.5\columnwidth]{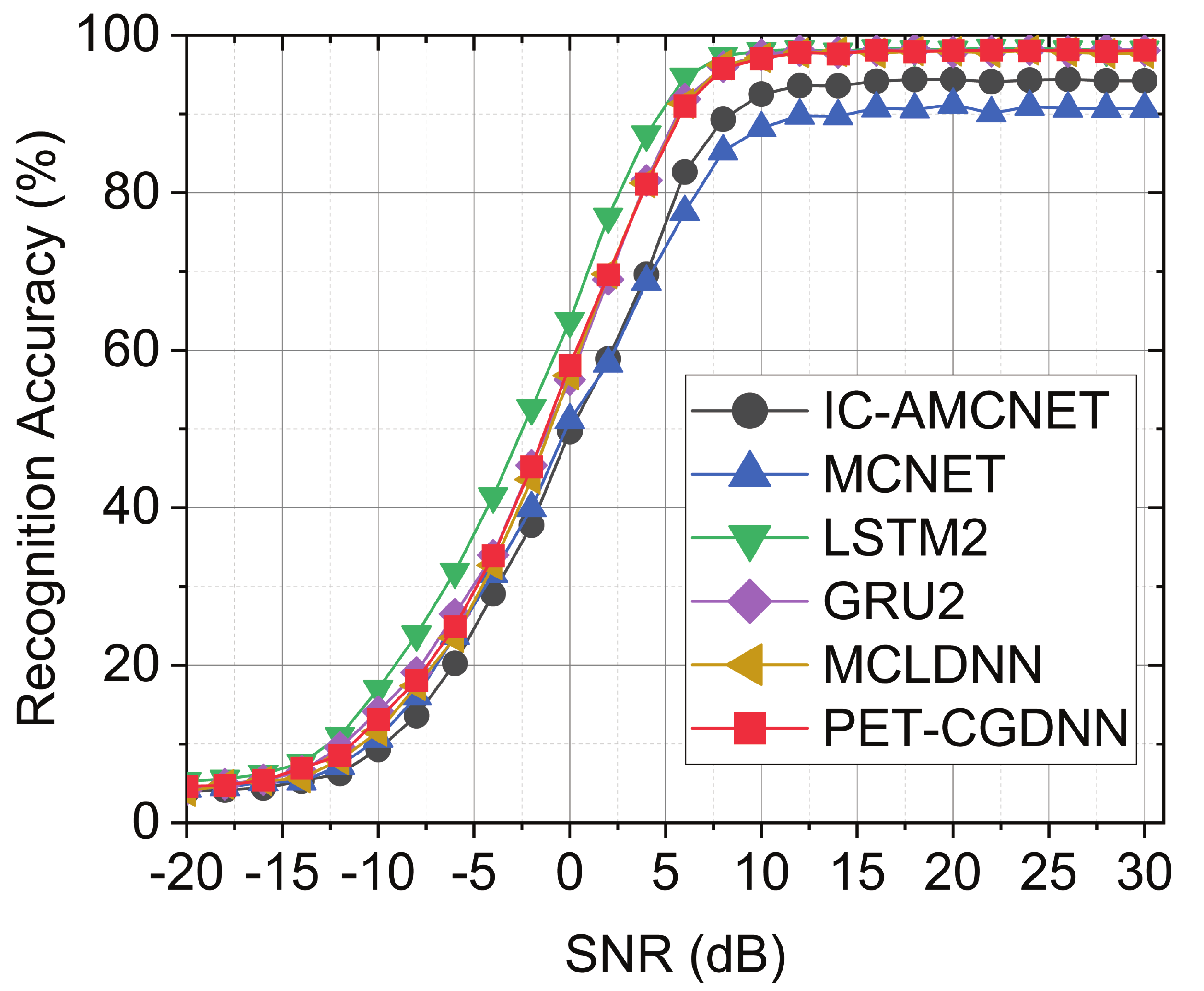}
\label{fig18}}
\subfloat[]{\includegraphics[width=0.5\columnwidth]{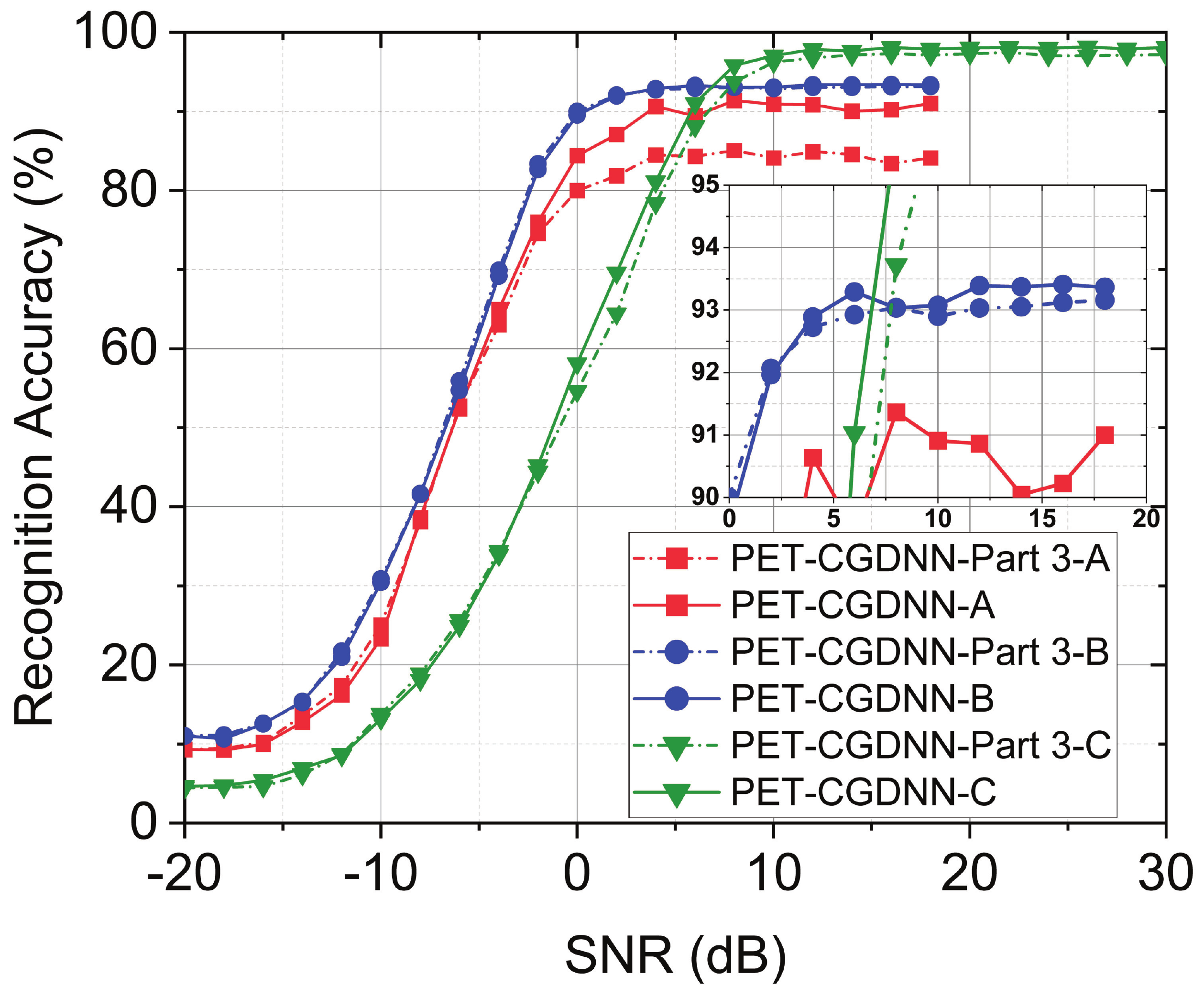}
\label{figcom}}
\caption{Recognition accuracy comparison on different datasets: (a) RML2016.10a, (b) RML2016.10b, (c) RML2018.01a, (d) ablation experiment comparison between PET-CGDNN and PET-CGDNN-Part 3 in Section \ref{Effectiveness Analysis} (A: RML2016.10a, B: RML2016.10b, C: RML2018.01a).} 
\label{fig4}
\end{figure}

Fig. \ref{fig4}\subref{fig16a} to Fig. \ref{fig4}\subref{fig18} provide detailed recognition accuracy against SNR on the three datasets. These figures demonstrate that our PET-CGDNN model maintains stable and consistent performance on all three datasets in comparison with the benchmark high accuracy models.
\subsection{Model Pruning}
\begin{figure}[htbp]
\centering
\subfloat[]{\includegraphics[width=0.5\columnwidth]{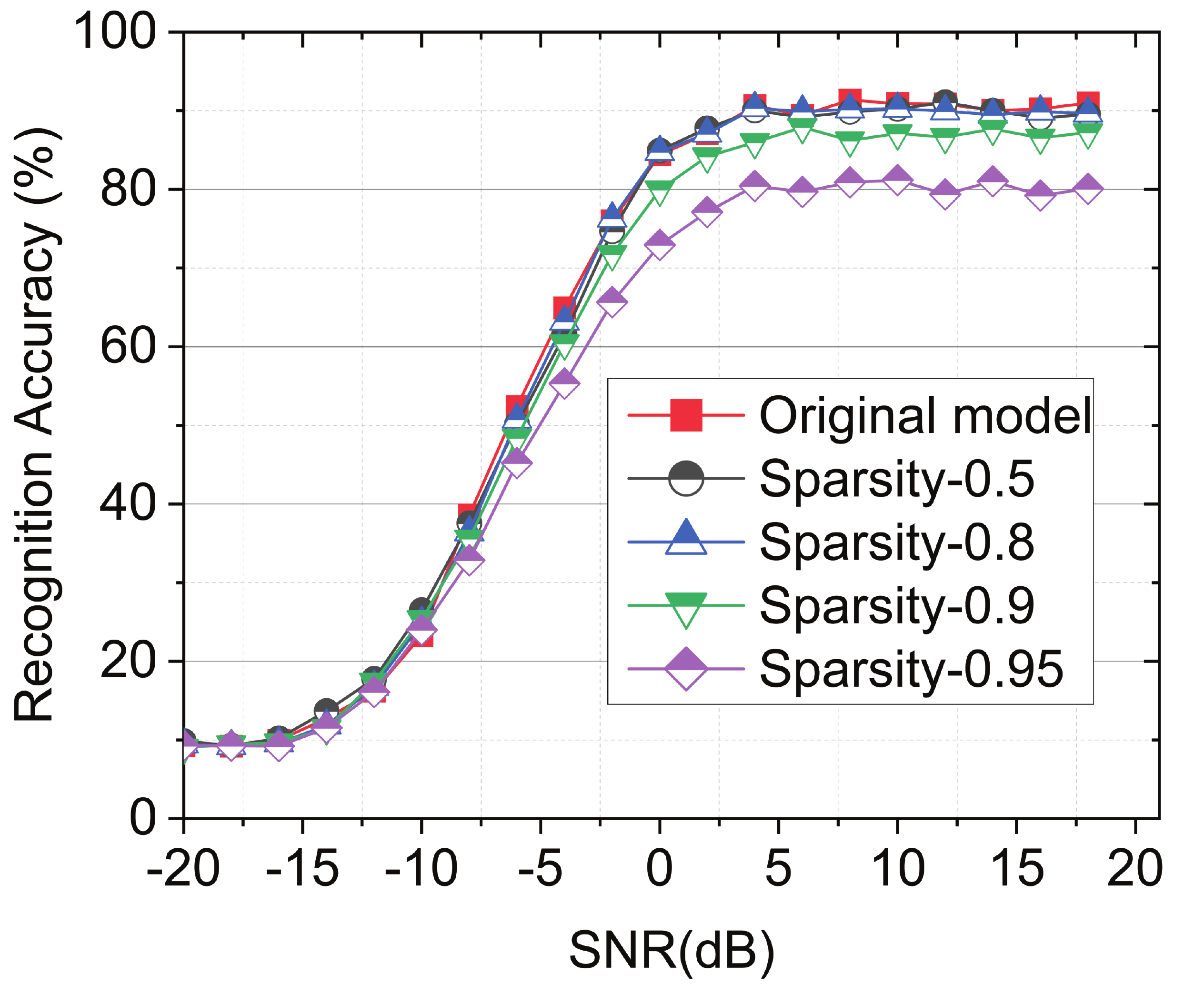}
\label{fig81}}
\subfloat[]{\includegraphics[width=0.5\columnwidth]{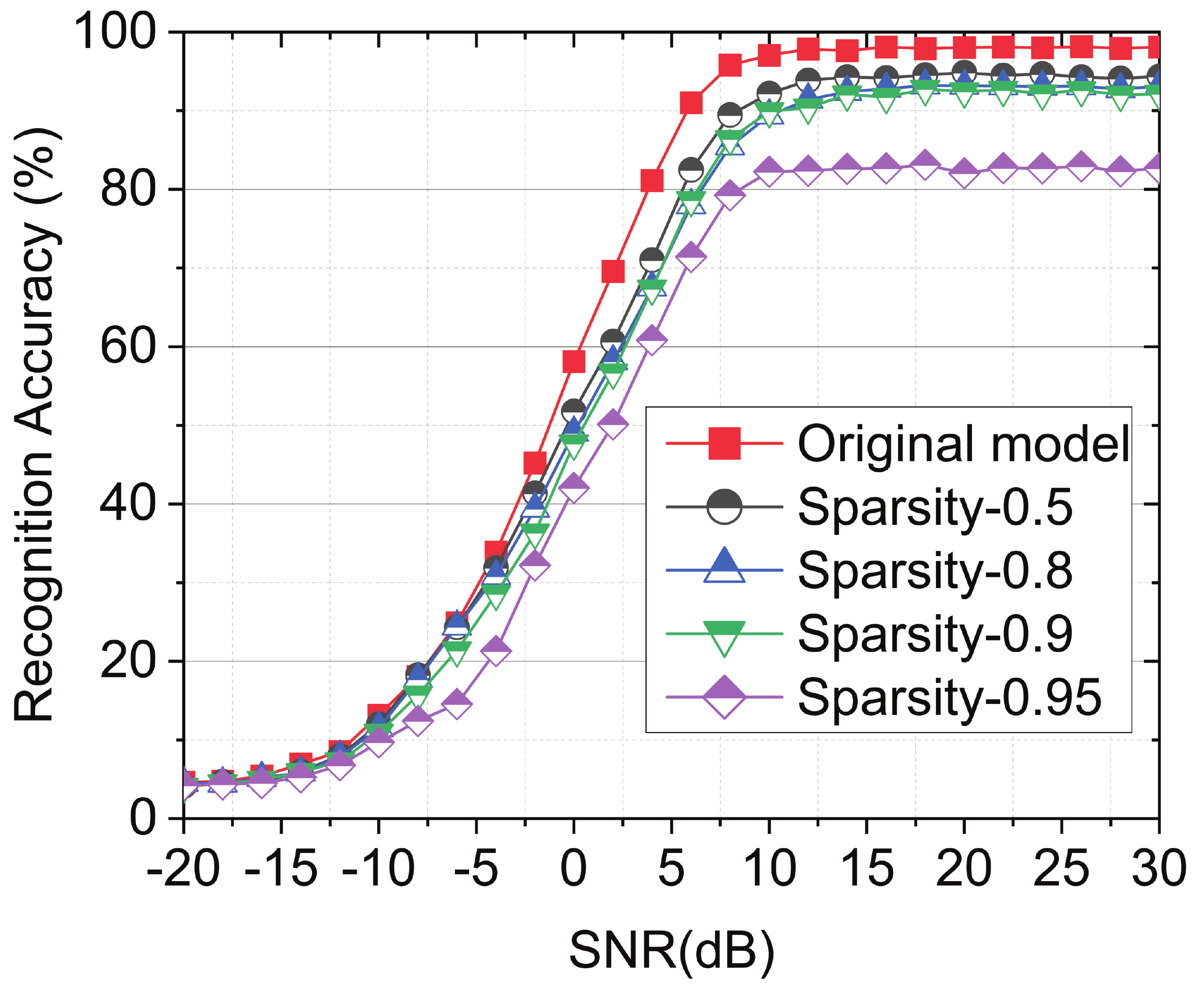}
\label{fig91}}
\caption{Model recognition accuracy after pruning: (a) RML2016.10a, (b) RML2018.01a.}
\label{fig8}
\end{figure}

We implement the pruning method in TensorFlow, which has been integrated into TensorFlow as a Keras-based pruning tool. In our experiment, the fine-tuning process has 5 epochs which are divided into 5,160 steps by setting the batch size to 128. The model is pruned and tested with different sparsity, while the size of the model is measured by the number of non-zero (NNZ) parameters. Table \ref{table2} and Fig. \ref{fig8} present the number of model parameters and the recognition accuracy after pruning. While the number of parameters is less than 15K with a sparsity of 0.8, the pruned model can maintain an accuracy above 90\%. The recognition accuracy of the pruned models remains stable on RML2016.10a and RML2016.10b when the sparsity is between 0 and 0.8, even though the model size is only 1/5 of the original model. The recognition accuracy of the pruned models decreases slightly on RML2018.01a, which indicates more connections are needed to fully extract information from the data with an input length of 1024. 

\begin{table}[htbp]
\centering
\caption{Model performance after pruning (A: RML2016.10a, B: RML2016.10b, C: RML2018.01a).}
\resizebox{0.4\textwidth}{!}{
\begin{tabular}{ccccc}
\hline
Sparsity    & Datasets&\begin{tabular}[c]{@{}c@{}}NNZ\\ parameters\end{tabular} & \begin{tabular}[c]{@{}c@{}}Highest\\ accuracy\end{tabular} 
 & \begin{tabular}[c]{@{}c@{}}Average\\ accuracy\end{tabular}\\ \hline
0 (Original) & \begin{tabular}[c]{@{}c@{}}A\\B\\C\end{tabular}& \begin{tabular}[c]{@{}c@{}}71K \\71K\\75K \end{tabular}& \begin{tabular}[c]{@{}c@{}}91.36\% \\93.41\%\\98.16\%\end{tabular}       &\begin{tabular}[c]{@{}c@{}}60.44\% \\ 63.82\%\\63.01\%\end{tabular}             \\\hline
0.5          & \begin{tabular}[c]{@{}c@{}}A\\B\\C\end{tabular}& \begin{tabular}[c]{@{}c@{}}35K \\35K\\37K \end{tabular} & \begin{tabular}[c]{@{}c@{}}91.09\%  \\93.28\% \\94.86\%\end{tabular}& \begin{tabular}[c]{@{}c@{}}60.43\% \\ 63.68\% \\59.49\%\end{tabular}   \\\hline
0.8          & \begin{tabular}[c]{@{}c@{}}A\\B\\C\end{tabular}& \begin{tabular}[c]{@{}c@{}}14K\\14K\\15K\end{tabular} &  \begin{tabular}[c]{@{}c@{}}90.36\%   \\ 93.26\% \\93.19\%\end{tabular}    &\begin{tabular}[c]{@{}c@{}}59.87\% \\ 63.56\% \\57.99\%\end{tabular}           \\\hline
0.9          & \begin{tabular}[c]{@{}c@{}}A\\B\\C\end{tabular}& \begin{tabular}[c]{@{}c@{}}7K\\7K\\7.4K\end{tabular}  &\begin{tabular}[c]{@{}c@{}}87.91\%   \\92.78\% \\92.74\%\end{tabular}      &\begin{tabular}[c]{@{}c@{}}55.61\% \\62.59\% \\57.21\%\end{tabular}           \\\hline
0.95         & \begin{tabular}[c]{@{}c@{}}A\\B\\C\end{tabular}& \begin{tabular}[c]{@{}c@{}}3.6K\\3.6K\\3.7K\end{tabular} & \begin{tabular}[c]{@{}c@{}}81.18\%  \\91.18\%\\83.14\%\end{tabular}     &\begin{tabular}[c]{@{}c@{}}49.53\% \\61.17\% \\51.06\%\end{tabular}               \\ \hline
\end{tabular}}
\label{table2}
\end{table}

\subsection{Effectiveness Analysis of PET-CGDNN}
\label{Effectiveness Analysis}
To further illustrate the effectiveness of the proposed model, we conduct an ablation experiment to compare the recognition accuracy of PET-CGDNN with the final module: PET-CGDNN-Part 3, which reflects the contribution of Part 1 and Part 2 to the recognition accuracy. It can be seen from Fig. \ref{fig4}\subref{figcom} that the model (PET-CGDNN-Part 3) without the parameter estimation and transformation module cannot achieve the equivalent recognition accuracy of PET-CGDNN in the high SNR range (above 0 dB). The average recognition accuracy and the best recognition accuracy of PET-CGDNN exhibit overall better performance, while the model size and time cost are almost the same. 
\begin{figure}[htbp]
\centering
\includegraphics[scale=.3]{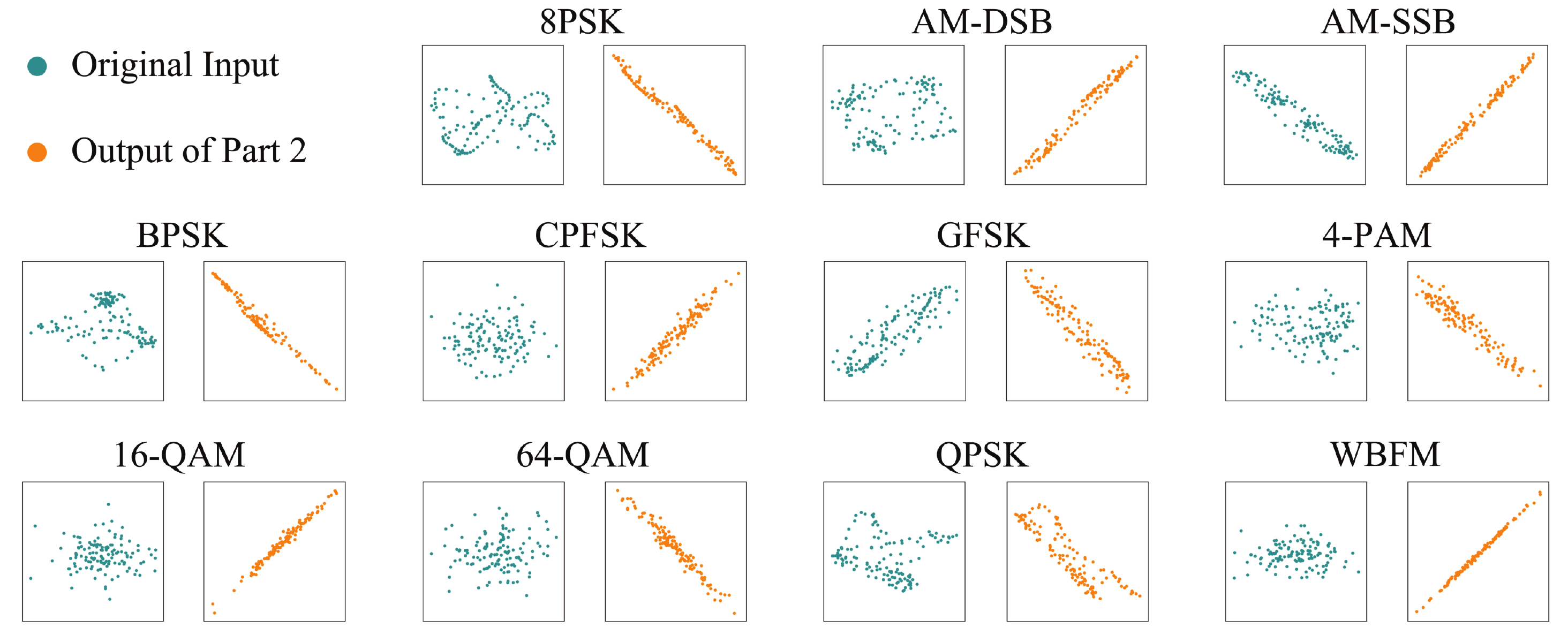}
\caption{The output of Part 2 and its comparison with the original input.}
\label{fig7}
\end{figure}
To further assess the functions of the parameter estimator and parameter transformer (Part 1 and Part 2) in PET-CGDNN, the output of Part 2 is visualized in the I/Q plane with a focus on the constellation distribution features rather than the value of these points (the origin of coordinates are different in each image). Fig. \ref{fig7} shows the output of Part 2 in comparison with the inputs to the model when the signals' SNR is +10 dB. It is clearly visible that the signals converge to tighter clusters after the two modules, which benefit the classification module and lead to the improvement of the overall recognition accuracy compared with PET-CGDNN-Part 3.


\section{Conclusion}
\label{section4}
In this letter, an efficient DL-AMR model, named PET-CGDNN, is proposed to achieve state-of-the-art performance. With the blessing of expert domain knowledge on phase offset estimation and compensation, the model enjoys the characteristics of lightweight, low-complexity, and high recognition accuracy, which also exhibits good stability on different datasets. Moreover, a pruning method has been applied to further reduce the model size, which exhibits the possibility to compress an AMR model, even if the model parameters are already very small. The efficient AMR model will have potentially wide application in the scenarios of massive machine-type communications and ultra-reliable and low latency communications in the future, which is in line with the development trend of future communication systems.

\ifCLASSOPTIONcaptionsoff
  \newpage
\fi

\bibliographystyle{IEEEtran}
\bibliography{IEEEexample}

\end{document}